\documentclass[showpacs,preprintnumbers,amsmath,amssymb,twocolumn]{revtex4}
\usepackage{amssymb}
\usepackage{amsfonts}

\usepackage{graphicx}
\usepackage{dcolumn}
\usepackage{bm}

\begin{document}

\title{A quantum model of dark energy}

\author{Chang-Yu Zhu$^1$ }

\affiliation{%
$^1$Department of Physics, Zhengzhou University, Zhengzhou, Henan
450052, China
}%

\author{Heng Fan$^{2}$
}
\affiliation{%
$^2$Institute of Physics, Chinese Academy of Sciences, Beijing
100190, China
}%
\date{\today}

\begin{abstract}
We propose a quantum model of dark energy. The proposed candidate
for dark energy is gluon field, as is well-known, gluons are the
elementary particles. We assume that gluons may not be completely
massless but have tiny masses, thus the gluon field can provide a
non-zero energy-momentum tensor. This model corresponds to
Einstein's cosmological constant which is one of the generally
accepted models for dark energy. Besides the gluon field, we also
discuss the properties of electroweak boson field and compare our
results with previous known results.
\end{abstract}

\maketitle

The observed evidences \cite{accelerate1,accelerate2} show that the
expansion of the universe is accelerating. The accelerated universe
is considered to be driven by the dark energy which accounts for
about $72\% $ of the total mass-energy of the universe \cite{WMAP}.
Recently, much effort has been put into studying this topic and
various models are proposed for dark energy. Most of those models
can be divided into as Einstein's cosmological constant or
quintessence, see Refs.\cite{Weinberg,RMP,IJMPD,ARA} for reviews and
related models. The quintessence models \cite{quinte} can provide an
explanation for dynamical process of cosmos such as for inflation in
the distant past. However, one of the basic questions of
quintessence models is that while a rolling scalar field is
necessary but how this quintessence field should fit in the
fundamental physics theory. The dark energy models of Einstein's
cosmological constant can provide an extremely simple, well-defined
mechanism for the acceleration. But a na\"{\i}ve calculations show
that the vacuum energy of the quantum field which takes the form of
cosmological constant is too big to be observationally acceptable
\cite{Weinberg}. It seems that dark energy point to physics beyond
the standard models of gravity or particle physics, and recently
various modifications of general relativity or quantum field
theories are proposed, see for example Ref. \cite{Zee}.

While we might consider to modify the standard theories of gravity
or quantum fields, we may still expect that a unified theory of
general relativity and quantum mechanics can provide an
interpretation of the phenomena of dark energy. In the present work,
we try to start from a unified theory which includes all four known
fundamental interactions of nature. With assumption that gluons,
which are elementary particles and are generally assumed to be
massless, may have tiny masses, we find that this quantum model of
dark energy fits well with observational facts while all
well-established results in Standard Model of particle physics and
general relativity remain to be unchanged. This model of dark energy
provides an origin of Einstein's cosmological constant.

Before going to details of the quantum model, let us briefly present
the main properties of dark energy. By the data from the Wilkinson
Microwave Anisotropy Probe (WMAP) \cite{WMAP} and other teams, the
universe is flat, homogeneous, and isotropic. The dark energy is
homogeneous, nearly independent of time and the density is very
small which is around $\rho _{\Lambda }=10^{-29}g/cm^{3}$. The dark
energy is not known to interact through any of the fundamental
forces except gravity. The gravitational effect of dark energy
approximates that of Einstein's cosmological constant with equation
of state parameter $w=-1$ within $14\% $ level. It has a strong
negative pressure which can explain the observed accelerating
universe. In this work, the quantum model of dark energy is based on
the theory in Ref. \cite{ZF}, here let's briefly review some results
in this theory.

{\it Brief introduction of a quantum theory with gravity}-- It is
well known that Dirac equation and Yang-Mills equation play the
central roles in quantum theory and Einstein equation is also the
central of general relativity. Dirac equation is generally written
as, $(i{\gamma }^{\alpha }{D}_{\alpha }-\hat {m})|\Psi \rangle =0.$
Yang-Mills equation can be written as,
\begin{eqnarray}
{D}_{\alpha }{F}_a^{\alpha \beta }={J}_a^{\beta }, \label{ym-eq}
\end{eqnarray}
where the generalized generally covariant derivative operator is
defined as, ${D}_{\alpha }=-i{\theta }_{\alpha }^{\mu }\otimes
{p}_{\mu }+\frac {i}{2}{\Gamma }_{\alpha }^{\rho \sigma }\otimes
{s}_{\rho \sigma }-i{A}_{\alpha }^a\otimes {T}_a$. We may notice
that in this operator the first term is the usual derivative
operator with spin vierbein ${\theta }_{\alpha }^{\mu }$, ${T}_a$ in
the third term is the gauge charge operator in the Standard Model of
particle physics with symmetry $U(1)\times SU(2)\times SU(3)$ which
are well accepted \cite{standard}, and the second term here
corresponds to the effect of gravity. The current density is defined
as,  ${J}_a^{\beta }= {j}_a^{\beta }+M_a^b{A}_b^{\beta }$, which is
conserved, ${D}_{\beta }{J}_a^{\beta }=0$. The matter-field current
takes the standard form as, $j^{\beta }_a(x)=\langle \Psi
|{\varepsilon }(x) {\gamma }^{\beta }{T}_a|\Psi \rangle $. $M_a^b$
are the entries of mass matrix for interaction bosons, we have
$M^1_1=0$ for massless photon in electromagnetic field,
$M^2_2=m_Z^2, M_3^3=M_4^4=m_W^2$ for electroweak bosons $Z_0,W_{\pm
}$ with masses $m_Z,m_W$, respectively, $M^j_j=m^2, j=5,6,\cdots
,12$, for 8 gluons in color interactions, each with mass $m$, other
entries are defined to be zeroes. Einstein equation as usual takes
the form, ${R}^{\alpha }_{\beta }-\frac {1}{2}\delta _{\beta
}^{\alpha }{R}=-8\pi G {T}^{\alpha }_{\beta }$, where $G$ is the
gravity constant, $R^{\alpha }_{\beta }$ is the Ricci tensor, $
{T}^{\alpha }_{\beta }$ is the energy-momentum tensor of all matters
and fields. Generally, the validity of equations of motion is based
on a proper Lagrangian. Here the combined and self-consistent form
of Dirac equation, Yang-Mills equation and Einstein equation can
lead to the energy-momentum conservation law presented as,
${D}_{\alpha }{T}^{\alpha }_{\beta }=0$. If no gravity is
considered, Dirac equation and Yang-Mills equation will reduce to
their standard forms as in Standard Model of particle physics. We
expect that those equations may be related with a Lagrangian. In
Standard Model of particle physics, gluons are assumed to be
massless. However, in this work, we assume that each gluon has mass
$m$. Apparently $m$ should not be large otherwise there might
already exist experimental evidences for massive gluons.

As in any Yang-Mills theory, the gauge curvature takes the form,
\begin{eqnarray}
{F}_{\alpha \beta }^{a}&=&{\theta }_{\alpha }^{\mu }\partial _{\mu
}{A}_{\beta }^a-{\theta }_{\beta }^{\mu }\partial _{\mu }{A}_{\alpha
}^a-iC_{bc}^a{A}_{\alpha }^b{A}_{\beta }^c -{f}_{\alpha \beta
}^{\gamma }{A}_{\gamma }^a, \label{gauge-curvature}
\end{eqnarray}
where $C_{bc}^a$ are the structure constants of gauge charge
operators ${T}_a$, and ${f}_{\alpha \beta }^{\gamma }$ is related
with spin vierbein, note that gauge fields ${A}_{\alpha }^a$ depend
on 4D coordinates ${x}_{\mu }$ which is standard. The
energy-momentum tensor of gauge fields takes the form
\begin{eqnarray}
{\tau }^{\alpha }_{\beta }={F}_a^{\alpha \rho }{F}^a_{\beta \rho
}-\frac {1}{4}\delta ^{\alpha }_{\beta }{F}_a^{\rho \sigma
}{F}^a_{\rho \sigma }+M^a_b{A}^{\alpha }_a{A}^b_{\beta }-\frac
{1}{2}\delta ^{\alpha }_{\beta }M^a_b{A}^{\rho }_a{A}^b_{\rho }.
\nonumber
\end{eqnarray}

{\it A gluon field as the dark energy}-- In order to find a dark
energy solution, let us make the following assumptions: (i) There is
no matters or matter particles, $|\Psi \rangle =0, j^a_{\alpha }=0$;
(ii) There is no gravity, $\Gamma ^{\rho \sigma }_{\alpha }=0$;
(iii) The Yang-Mills fields are constant, $\partial _{\mu }
A^a_{\alpha }=0$. The observation facts show that the dark energy is
generally not any matter particles including neutrinos, it seems not
photon, and actually there is not at all any matters. So we make
assumption (i). The assumption (ii) is based on the observational
facts that universe is flat, isotropic and homogeneous. The
assumption (iii) means that momentum of the field is zero since of
the constant gauge fields and correspondingly it does not change in
4D coordinate space. That means this field is invariant in 4D
space-time and it will not cause any energy excitation, it is like
the vacuum state. If we believe that dark energy should correspond
to some elementary particles, since of the above reasons, a field of
gluons seems the most possible choice. So we assume that (iv) the
dark energy is the field of gluons.

Recall the definition in Eq. (\ref{gauge-curvature}), we find the
gauge curvature now takes the form
\begin{eqnarray}
F^a_{\alpha \beta }=-iC^a_{bc}A^b_{\alpha }A^c_{\beta }. \nonumber
\end{eqnarray}
Substitute the gauge curvature into Yang-Mills equation in
(\ref{ym-eq}), we have,
\begin{eqnarray}
D^{\alpha }F^a_{\alpha \beta }=M^a_bA^b_{\beta }, \nonumber
\end{eqnarray}
so we have a relation, $-iC^a_{bc}A^{b,\alpha }F^c_{\alpha \beta
}=M^a_bA^b_{\beta }$.  Thus the equation need to be solved is,
\begin{eqnarray}
-C^a_{bc}C^c_{de}A^{b,\alpha }A^d_{\alpha }A^e_{\beta }=M^a_eA^e_{
\beta }. \nonumber
\end{eqnarray}
One solution of this equation is $A^e_{\beta }=0$, it is trivial and
we do not discuss it. Next, we shall consider the solution,
\begin{eqnarray}
-C^a_{bc}C^c_{de}A^{b,\alpha }A^d_{\alpha }=M^a_e.
\label{dark-solution}
\end{eqnarray}
By considering the definition of the curvature and the symmetries of
the structure constants, we can find that,
\begin{eqnarray}
F^{\alpha \rho }_aF^a_{\beta \rho }&=&-M^b_dA^{\alpha }_bA^d_{\beta
}, \nonumber
\end{eqnarray}
where the solution (\ref{dark-solution}) is already used, also we
have,
\begin{eqnarray}
F^{\rho \sigma }_{a}F^a_{\rho \sigma }=-M^b_dA^{\rho }_bA^d_{\rho }.
\nonumber
\end{eqnarray}
Substituting those results to the energy-momentum tensor of gauge
fields, we can find that,
\begin{eqnarray}
\tau ^{\alpha }_{\beta }&=&-\frac {1}{4}\delta ^{\alpha }_{\beta
}M^a_bA^{\rho }_aA^b_{\rho }. \label{dark-all-choice}
\end{eqnarray}
The explicit form of this energy-momentum tensor depends on the
solution of equation (\ref{dark-solution}) if the exact mass matrix
$M^a_e$ is given.

Next, we consider about the case of elementary particles, 8 types of
gluon. For this case, the mass matrix of gluons takes the form,
\begin{eqnarray}
M^a_e=m^2\delta ^a_e, \label{mass-gluon}
\end{eqnarray}
where $m$ is the mass of each gluon, $a,e=5,6,\cdots ,12$. We then
consider the solution of Yang-Mills equation (\ref{dark-solution}),
we find,
\begin{eqnarray}
m^2&=&-g_3^2A^{\alpha }_dA^d_{\alpha }, \nonumber
\end{eqnarray}
where $g_3\approx 1.22$ is the coupling constant in the Standard
Model of particle physics. Consider also the form of mass matrix of
gluons, the energy-momentum tensor of gluon field is,
\begin{eqnarray}
\tau ^{\alpha }_{\beta }=\delta ^{\alpha }_{\beta }\frac
{m^4}{4g_3^2}. \nonumber
\end{eqnarray}
Recall that the energy-momentum tensor of matter-field is zero, the
total energy-momentum tensor only has contribution of the gluon
field, thus
\begin{eqnarray}
T_{\alpha ,\beta }=\eta _{\alpha \beta }\frac {m^4}{4g_3^2},
\label{dark-en-mo}
\end{eqnarray}
where $\eta _{\alpha \beta }$ is the Minkowski metric with diagonal
entries as $\eta _{\alpha \beta }={\rm diag.}(1,-1,-1,-1)$. Please
note that no boundary condition is used to obtain this solution, and
it can be assumed to be valid for universe. Also this is a unique
solution. It is now clear that the energy-momentum tensor here
(\ref{dark-en-mo}) corresponds to Einstein's cosmological constant.
So we conclude that the density of the dark energy $\rho _{\Lambda
}$ is,
\begin{eqnarray}
\rho _{\Lambda }=\frac {m^4}{4g_3^2}.
\end{eqnarray}
Consider that the density of the dark energy $\rho _{\Lambda
}\approx 10^{-29}g/cm^{3}$, we can estimate that the mass of each
gluon is around $10^{-3}eV\sim 10^{-2}eV$. It is theoretically
accepted and experimentally confirmed that the masses of gluons are
zeroes, here we can see that the estimated mass of each gluon is
very small thus it should not have detectable effects on present
experiments. As is well-known, the gluon field only interacts
through color (strong) interactions, also note that there is the
confinement of gluons due to color confinement, so a free gluon is
impossible or hard to be detected directly in experiments. Actually
since our assumption (iii), this gluon field will always be
invariant. Thus it only provides the effect of the Einstein's
cosmological constant, note that it has the origin from the quantum
theory and is independent of gravity. The result that dark energy is
the gluon field does not change standard results of quantum theory
and general relativity.

The elementary particle gluon was discovered by several experiments
almost at the same time in electron-positron annihilation process
which was indicated by the three-jet events \cite{gluon}.
Theoretically, gluons are assumed to be massless, the present
experiments put the upper bound of the masses of gluon at around
$MeV$ level \cite{jpg}. In this work, we estimate that the mass of
each gluon is around $10^{-3}eV\sim 10^{-2}eV$ which is still far
from $MeV$ level. It seems that a direct observation of the massive
gluons in laboratory might still be difficult. However, recent
experiments can already make the observation of single top quark
production \cite{topquark1,topquark2}. As we know since of the color
confinement, a single quark is difficult to be observed. We expect
that this experimental improvements can make it possible to observe
free gluons. Moreover, the additional properties of dark energy
observed could provide us more information  to test whether it is
really related with the gluon field.

{\it An electroweak interactions boson field and comparison with
known results}-- In case the color interactions field, the gluon
field, may be considered as the candidate for dark energy. It is
naturally to ask whether the electroweak interactions boson field
can play a role in universe.

Let us still keep assumptions (i), (ii) and (iii) as presented
above, now the additional assumption is that (iv') we consider about
the electroweak boson field. As is well-known, the elementary
particles of electroweak bosons are $Z_0, W_{\pm }$ which have
masses $m_Z\approx 91GeV, m_W\approx 80GeV$. So as we mentioned, the
corresponding elements of the mass matrix are $M^2_2=m_Z^2,
M_3^3=M_4^4=m_W^2$, the related part of mass matrix does not have a
symmetry similar as for color gluons presented in
(\ref{mass-gluon}). However, as we know, the isospin of electroweak
bosons satisfies a $SU(2)$ symmetry and also the mass of photon is
zero and the corresponding hypercharge commutes with isospin
charges, so mass matrix of $W_{\pm },Z_0$ also has a $SU(2)$
symmetry. By proper definition, we may find that $\tilde
{M}^i_i=m_W^2, i=2,3,4$. Now let us make the calculations similar as
for gluon field, we find the energy-momentum tensor of electroweak
boson field takes the form,
\begin{eqnarray}
T_{\alpha ,\beta }^{e-w}=\eta _{\alpha \beta }\frac {m^4_W}{4g_2^2},
\label{electro-weak-mo}
\end{eqnarray}
where $g_2\approx 0.65 $ is the electroweak coupling constant in
Standard Model. So the mass-energy density of electroweak boson
field is,
\begin{eqnarray}
\rho _{e-w}=\frac {m^4_W}{4g_2^2}. \label{weak-em}
\end{eqnarray}
We thus can estimate, $\rho _{e-w}\approx 2.4\times 10^{7} GeV^4$.
The magnitude of this result arouse us the previous known results
concerning about the vacuum energy to be the cosmological constant
in Ref.\cite{Weinberg} by Weinberg and others, see also
Ref.\cite{cosmos} by Dreitlein.

To confirm that the results in this work are correct, let's briefly
review the results in Ref.\cite{cosmos}, see also Ref.
\cite{Weinberg}. For convenience we follow the main notations of
\cite{cosmos}. Suppose the system is the electroweak model with
$U(1)\times SU(2)$ gauge symmetry, $V(\varphi )$ is the potential
term in the Higgs-field Lagrangian. The only nonvanishing part of
the energy-momentum tensor is, $\langle T_{\alpha \beta }\rangle
=\eta _{\alpha \beta }\langle V(\varphi )\rangle$, which is related
with the vacuum energy, where
\begin{eqnarray}
V(\varphi )=\mu ^2\varphi ^{\dagger }\phi +\lambda (\varphi
^{\dagger }\varphi )^2.
\end{eqnarray}
By considering the physical parameters, we can find that
\begin{eqnarray}
\langle V(\varphi )\rangle =\frac {m_{\varphi }^2m_W^2}{2g_2^2}.
\end{eqnarray}
The system considered in Ref.\cite{cosmos} is actually a coupled
Higgs scalar boson field with mass $m_{\varphi }$ and the
electroweak boson field. If we let this Higgs scalar boson also be
the weak boson $m_{\varphi }=m_W$ which is what we considered in
this work, we find that
\begin{eqnarray}
\langle V(\phi )\rangle =\frac {m_W^4}{2g_2^2}. \label{higgs-em}
\end{eqnarray}
The result in this work presented as (\ref{weak-em}) is exactly the
same as the result (\ref{higgs-em}) presented in Ref.\cite{cosmos}
up to a factor $2$ due to different definitions. The fact that two
different methods lead to the same conclusion confirms that the
calculations of this work are correct and the method is reasonable.

{\it Discussions and conclusions}-- One may notice that $\rho
_{e-w}$ is too large to be considered as the dark energy. However,
if there exists such a matter constituted only by electroweak
bosons, the strong negative pressure will cause a big explosion,
this might be related with the universe model of Big Bang. The
density $\rho _{e-w}$ which is around $10^{25}g/cm^3$ is also very
huge, possibly it is suggestive to consider it as the kernel of
black hole.

In conclusion, if we believe that dark energy should fit in the
standard particle physics, the proposed gluon field is likely to be
the candidate. The massive gluon field in the Standard Model of
particle physics generally means that Higgs mechanism should be used
to create this mass which also implies that additional scalar fields
should be introduced. In case that those scalar fields are
dynamical, the proposed quantum model of dark energy in this work
should correspond to the quintessence model. From this point of
view, we can roughly say that the present work is a combination of
models of cosmological constant and quintessence.

Let's remark some properties of gluons. The interaction of gluons is
short-ranged, the gluons are color charged. Gluons are assumed to be
the mediation particles among quarks for strong interactions and the
gluon can interact with itself. There is color confinement in
particle physics due to the $SU(3)$ symmetry thus single quark and
free gluons are difficult to be observed. However experiments have
already succeeded to make single top quark production
\cite{topquark1,topquark2}. For case of gluon, actually, it is a
long history of assumption theoretically and experimentally that
pure gluon field without quark, which is named as glueballs, may
exist \cite{glueball,bak}. We may expect that the experiments can
finally test whether a massive or a massless gluon field exist or
not.

By comparing our results with previous known results in electroweak
interactions field case, we find that the constant solution of the
Yang-Mills equation in our formalism corresponds to the vacuum
energy of a quantum field by known Higgs Lagrangian method. One
might consider some other choices for dark energy, but the observed
facts exclude any known matter particles being the dark energy, the
only left choices known in elementary particles are interactions
bosons such as photon, electroweak bosons or gluons. We analyzed
those possibilities and showed that a gluon field can be the
candidate. One advantages of this theory is that while a massive
gluon field can be proposed, we can still keep the standard results
of quantum theory and general relativity.

{\it Acknowledgements--}This work is supported by grants of National
Natural Science Foundation of China (NSFC) Nos.(10674162,10974247),
``973'' program (2010CB922904) of Ministry of Science and Technology
(MOST), China, and Hundred-Talent Project of Chinese Academy of
Sciences (CAS), China.


\begin{thebibliography}{99}




\bibitem{accelerate1} A. G. Riess {\it et al.}
Astronomical J. {\bf 116}, 1009 (1998).

\bibitem{accelerate2}S. Perlmutter {\it et al.}
Astronomical J.
{\bf 517}, 565 (1999).

\bibitem{WMAP}E. Komatsu {\it et al.}
Astrophys. J. Suppl. {\bf 180}, 330 (2009).

\bibitem{Weinberg}S. Weinberg, Rev. Mod. Phys. {\bf 61}, 1 (1989).

\bibitem{RMP}P. J. E. Peebles, and R. Bharat,
Rev. Mod. Phys. {\bf 75}, 559 (2003).

\bibitem{IJMPD}E. J. Copeland,  M. Sami,  and S. Tsujikawa,
Int. J. Mod. Phys. D {\bf 15}, 1753 (2006).

\bibitem{ARA}J. A. Frieman,  M. S. Turner, and D. Huterer,
Ann. Rev. Astron. and Astrop. {\bf 46}, 385 (2008).


\bibitem{quinte}P. J. E. Peebles and A. Vilenkin, Phys. Rev.
D {\bf 59}, 063505 (1999).

\bibitem{Zee}R. A. Porto, and A. Zee, eprint arXiv:0910.3726.

\bibitem{ZF} C. Y. Zhu, and  H. Fan,
eprint arXiv:0911.1402.

\bibitem{standard}see, for example, T. P. Cheng, \& L. F. Li,
``Gauge theory of elementary particle physics, '' Oxford University
Press (2006); E. Golowich, J. F. Donoghue and B. R. Holstein,
``Dynamics of the Standard Model,'' Cambridge University Press,
Cambridge (1992); S. Weinberg, ``The quantum theory of
fields,''Cambridge University Press, Cambridge (1996);  S. F.
Novaes, eprint arXiv:hep-ph/0001283.

\bibitem{gluon}JADE collaboration, Phys. Lett. B {\bf 91}, 142 (1980);
TASSO collaboration, Phys. Lett. B{\bf 86}, 243 (1979);
MARK collaboration, Phys. Rev. Lett. {\bf 43}, 830 (1979); PLUTO
collaboration, Phys. Lett. B {\bf 86}, 418 (1979).

\bibitem{jpg}W. M. Yao {\it et al.} (Particle Data Group), J. Phys.
G {\bf 33}, 1 (2006).

\bibitem{topquark1}T. Aaltonen {\it et al.} (CDF collaboration),
Phys. Rev. Lett. {\bf 103}, 092002 (2009).

\bibitem{topquark2}V. M. Abazov {\it et al.} (D0 collaboration),
Phys. Rev. Lett. {\bf 103}, 092002 (2009).

\bibitem{cosmos}J. Dreitlein, Phys. Rev. Lett. {\bf 33}, 1243
(1974).

\bibitem{glueball}R. L. Jaffe and K. Johnson, Phys. Lett. B {\bf
60}, 201 (1976).

\bibitem{bak}P. van Baal and A. S. Kronfeld, Nucl. Phys. (Proc.
Suppl.) {\bf 9}, 227 (1989).
\end{thebibliography}
\end{document}